\documentclass[sigconf,screen]{acmart}

\makeatletter
\def\@ACM@checkaffil{
    \if@ACM@instpresent\else
    \ClassWarningNoLine{\@classname}{No institution present for an affiliation}%
    \fi
    \if@ACM@citypresent\else
    \ClassWarningNoLine{\@classname}{No city present for an affiliation}%
    \fi
    \if@ACM@countrypresent\else
        \ClassWarningNoLine{\@classname}{No country present for an affiliation}%
    \fi
}
\makeatother

\usepackage{amsmath}

\usepackage{amssymb,amsfonts}
\usepackage{ragged2e}
\usepackage{algorithmic}
\usepackage{graphicx}
\usepackage{textcomp}
\usepackage{enumitem}
\usepackage{xcolor}
\usepackage[many]{tcolorbox}
\usepackage{soul}
\usepackage{booktabs}
\usepackage{caption}
\usepackage{breakurl}
\usepackage{hyperref}
\usepackage{multirow}
\usepackage{svg}
\usepackage{lipsum}
\usepackage[english]{babel}
\usepackage[autostyle]{csquotes}
\usepackage{pdfpages}

\newcommand\blfootnote[1]{%
  \begingroup
  \renewcommand\thefootnote{}\footnote{#1}%
  \addtocounter{footnote}{-1}%
  \endgroup
}

\AtBeginDocument{%
  \providecommand\BibTeX{{%
    \normalfont B\kern-0.5em{\scshape i\kern-0.25em b}\kern-0.8em\TeX}}}

\usepackage{soul}

\title[Synthesizing Speech Test Cases with Text-to-Speech? ...]{Synthesizing Speech Test Cases with Text-to-Speech? \\ An Empirical Study on the False Alarms in Automated Speech Recognition Testing}

\author{Julia Kaiwen Lau}
\email{jlau0019@student.monash.edu}
\affiliation{%
\institution{School of Information Technology, Monash University Malaysia}
\city{Subang Jaya}
\country{Malaysia}}
\author{Kelvin Kai Wen Kong}
\email{kkon0010@student.monash.edu}
\affiliation{%
\institution{School of Information Technology, Monash University Malaysia}
\city{Subang Jaya}
\country{Malaysia}}
\author{Julian Hao Yong}
\email{jyon0017@student.monash.edu}
\affiliation{%
\institution{School of Information Technology, Monash University Malaysia}
\city{Subang Jaya}
\country{Malaysia}}
\author{Per Hoong Tan}
\email{ptan0021@student.monash.edu}
\affiliation{%
\institution{School of Information Technology, Monash University Malaysia}
\city{Subang Jaya}
\country{Malaysia}}

\author{Zhou Yang}
\email{zyang@smu.edu.sg}
\affiliation{%
\institution{School of Computing and Information Systems, Singapore Management University}
\city{Singapore}
\country{Singapore}}

\author{Zi Qian Yong}
\email{zyon0005@student.monash.edu}
\affiliation{%
\institution{School of Information Technology, Monash University Malaysia}
\city{Subang Jaya}
\country{Malaysia}}
\author{Joshua Chern Wey Low}
\email{clow0007@student.monash.edu}
\affiliation{%
\institution{School of Information Technology, Monash University Malaysia}
\city{Subang Jaya}
\country{Malaysia}}
\author{Chun Yong Chong}
\email{chong.chunyong@monash.edu}
\affiliation{%
\institution{School of Information Technology, Monash University Malaysia}
\city{Subang Jaya}
\country{Malaysia}}
\author{Mei Kuan Lim}
\email{lim.meikuan@monash.edu}
\affiliation{%
\institution{School of Information Technology, Monash University Malaysia}
\city{Subang Jaya}
\country{Malaysia}}
\author{David Lo}
\email{davidlo@smu.edu.sg}
\affiliation{%
\institution{School of Computing and Information Systems, Singapore Management University}
\city{Singapore}
\country{Singapore}}

\renewcommand{\shortauthors}{Lau et al.}

\begin{document}

\begin{abstract}
Recent\blfootnote{\textsuperscript{$\ddag$}Zhou Yang is the corresponding author.} studies have proposed the use of Text-To-Speech (TTS) systems to automatically synthesise speech test cases on a scale and uncover a large number of failures in ASR systems. 
However, the failures uncovered by synthetic test cases may not reflect the actual performance of an ASR system when it transcribes human audio, which we refer to as \textit{false alarms}. 
Given a failed test case synthesised from TTS systems, which consists of TTS-generated audio and the corresponding ground truth text, we feed the human audio stating the same text to an ASR system. 
If human audio can be correctly transcribed, an instance of a \textit{false alarm} is detected.

In this study, we investigate false alarm occurrences in five popular ASR systems using synthetic audio generated from four TTS systems and human audio obtained from two commonly used datasets.
Our results show that the least number of false alarms is identified when testing Deepspeech, and the number of false alarms is the highest when testing Wav2vec2. On average, false alarm rates range from 21\% to 34\% in all five ASR systems. 
Among the TTS systems used, Google TTS produces the least number of false alarms (17\%), and Espeak TTS produces the highest number of false alarms (32\%) among the four TTS systems. 
Additionally, we build a false alarm estimator that flags potential false alarms, which achieves promising results: a precision of 98.3\%, a recall of 96.4\%, an accuracy of 98.5\%, and an F1 score of 97.3\%. 
Our study provides insight into the appropriate selection of TTS systems to generate high-quality speech to test ASR systems.
Additionally, a false alarm estimator can be a way to minimise the impact of false alarms and help developers choose suitable test inputs when evaluating ASR systems. 
The source code used in this paper is publicly available on GitHub at \url{https://github.com/julianyonghao/FAinASRtest}.
\end{abstract}



\maketitle

\section{Introduction}
Automatic Speech Recognition (ASR) allows users to interact with devices using their voices. 
ASR systems are prevalent in our daily lives~\cite{yu2016automatic}. 
Popular applications of ASR systems include mobile virtual assistants such as Siri\footnote{\url{https://www.apple.com/siri/}} and Alexa.\footnote{\url{https://developer.amazon.com/en-GB/alexa}}
As ASR systems become increasingly popular, it is crucial to ensure that ASR systems are capable of correctly recognising speech. 
To test an ASR system, an audio input and its corresponding transcription are required, which we call a \textit{speech test case}.
Collecting speech test cases manually, where recording and transcribing audio inputs are largely human-based, is rather laborious and costly. 
As a result, it is critical to automate the process of generating speech test cases on a scale~\cite{jain2019automated}.

Recent studies propose automated testing methods for ASR systems by using Text-To-Speech (TTS) systems, such as CrossASR~\cite{asyrofi2020crossasr}. 
TTS is a technology that accepts text as input and produces audio as output~\cite{olive1985text}. 
Advances in deep neural networks have made TTS systems more powerful, accessible, and inexpensive to use.
CrossASR uses TTS systems to automatically generate test cases for ASR systems. 
Instead of manually recording and transcribing audio, a text (ground truth) is fed into the TTS to produce TTS-generated audio, which in return is used to test the ASR system. 
Without the need to manually record and transcribe audio, ASR system developers can quickly scale up and automate the testing process by feeding a large text corpus into the TTS to produce test cases.

However, ASR systems are mainly trained and tested to recognise human audio in practical environments.
One question of using TTS-generated audio to test ASR systems naturally arises: \textit{Do failed test cases uncovered by TTS-generated audio reveal real faults when the ASR system transcribes human audio?}
An ASR system may fail to transcribe TTS-generated audio, but can correctly transcribe human audio stating the same text. 
Such a synthetic failed test case does not suggest an issue with the ASR as it is able to correctly recognise human audio. 
Instead, it suggests an issue with test case generation using TTS, which we refer to as \textit{false alarms}.

False alarms can have a significant impact on the evaluation and improvement of ASR systems. 
During the testing process of a software system, it is usually indicative of a code defect when a test case fails. 
However, not all test failures imply that there are code defects; the failure may be due to faulty test cases~\cite{herzig2015empirically}. Similar cases can occur when testing ASR systems with TTS-generated audio. 
When false alarms occur, developers need to troubleshoot the ASR systems to identify the issue, not knowing that the issue is actually caused by the test case using TTS-generated audio. Such false alarms provide less information on the improvement of ASR systems and impede the testing process. 
Although some tools have been proposed to identify false alarms in conventional software testing~\cite{kang2022detecting, kharkar2022learning, yoon2014reducing, hanam2014finding}, there is still a gap in false alarm identification for ASR testing as many of these tools target false alarms caused by static analysis tools, which are used to identify potential bugs and security vulnerabilities in source codes.

There are a variety of factors that can cause false alarms. One of the factors is that, since the main function of an ASR system is to transcribe human audio in operational environments, many state-of-the-art ASR systems are trained with human audio rather than TTS-generated audio. 
For example, Deepspeech~\cite{hannun2014deep} is trained with 5,000 hours of audio by 9,600 speakers. Deepspeech2~\cite{amodei2016deep} is also trained with large amounts of labelled data consisting of human audio with the corresponding transcriptions. 
Although modern TTS systems try to mimic human speech, they can potentially lack certain aspects of human speech, e.g., variations in pitch, intensity, duration, and speech sections~\cite{taake2009comparison}. 
Additionally, TTS-generated audio generally lacks natural phonetic variability~\cite{roring2007age} and lacks redundant acoustic cues present in natural speech~\cite{winters2004perception}. 
The above factors make the TTS-generated audio follow a data distribution that is different from that of an ASR system's training data. 
Deep neural network (DNN)-based systems are known to perform poorly in out-of-distribution data. 
Thus, the failure identified using TTS-generated audio may not reflect the actual performance of an ASR system.

The primary objective of this research is to perform an empirical assessment of false alarms in automated speech recognition (ASR) testing and develop a false alarm estimator. 
The estimator serves as a tool to anticipate potential false alarms. 
To achieve this, we conducted experiments using five distinct ASR systems and four TTS systems.
The ASR systems are Deepspeech~\cite{hannun2014deep}, Deepspeech2~\cite{amodei2016deep}, Vosk~\cite{vosk}, Wav2letter++~\cite{collobert2016wav2letter}, and Wav2vec2~\cite{baevski2020wav2vec}. 
The TTS systems include Google TTS~\cite{googletts}, Espeak~\cite{espeaktts},
Festival~\cite{festival},and GlowTTS~\cite{glowtts}. 
We use the LJ Speech Dataset~\cite{ljspeech17} and the LibriSpeech Dataset~\cite{panayotov2015librispeech}, which consist of 9,925 and 12,000 pairs of text and human audio, respectively.

Of all test cases synthesised using TTS systems, 20.79\%, 32.25\%, 12.71\%, 25.82\%, and 34.16\% of failed test cases are identified as false alarms when tested with Deepspeech, Deepspeech2, Vosk, Wav2letter, and Wav2vec2, respectively.
While Vosk has produced the least number of false alarms, its performance differs when tested on different datasets.
On the LJ Speech Dataset, Vosk achieves a low Word Error Rate (WER) value of 0.1 when transcribing human audio.
However, Vosk performs poorly on the human audio from the LibriSpeech Dataset, which indicates that the ASR performs in an inconsistent manner.
DeepSpeech, on the other hand, performs consistently on both LJ Speech and LibriSpeech Datasets and yields a reasonably low number of false alarms. The highest number of false alarms is identified with the use of Wav2vec2.

Among all the Text-To-Speech (TTS) systems assessed, it is observed that the ASR systems that use Google TTS and GlowTTS exhibit the fewest instances of false alarms. This finding suggests that both TTS systems are proficient at generating superior quality speech test cases.
When manually reviewing false alarms, we find that a substantial number of them are attributed to the generation of unclear speech audio by the TTS systems.
Moreover, the false alarm estimator we propose, based on Recurrent Neural Networks (RNN), is capable of estimating the potential occurrences of false alarms.


The contributions of our paper are as follows: 
\begin{itemize}[leftmargin=*]
    \item We have evaluated the performance difference in ASR systems when transcribing TTS-generated audio and human audio and showed that ASR systems are generally better in transcribing human audio. 
    \item We identify a large number of false alarms when testing ASR systems with TTS-generated audio, raising awareness to developers who use TTS systems to test ASR systems. 
    \item Among all the TTS systems used, the ASR systems can effectively transcribe audio generated using Google TTS and GlowTTS. Developers can consider selecting Google TTS and GlowTTS to generate high-quality speech test cases to test ASR systems. 
    \item We propose a false alarm estimator that flags possible false alarms when ASR is tested with TTS-generated audio. 
\end{itemize}
 
\noindent
\textbf{Paper Structure:} Section \ref{Background} provides an overview of ASR systems, ASR testing, and CrossASR. In Section \ref{Methodology}, we present our approach to identify false alarms and train a false alarm estimator. 
Section \ref{Experiments} outlines our datasets and experimental settings. Our research questions and results are addressed in Section \ref{Results}, followed by a discussion of the findings in Section \ref{Discussion}. Additionally, we explore related work in Section \ref{Related Work}, address threats to validity in Section \ref{Threats}, and conclude our work and discuss future research in Section \ref{Conclusion}.

\section{Background} \label{Background}

\subsection{Automated Speech Recognition Systems}
Automated Speech Recognition (ASR) systems convert audio (input) into text to obtain textual information on the given speech ~\cite{yu2016automatic, ji2022asrtest}. 
The underlying architecture of an ASR primarily involves the process of accepting an input speech sequence $S=\{s_1, s_2, ...., s_N\}$ and recognising them as textual token sequence $T=\{t_1, t_2, ...., t_M\}$, which usually is in the form of phonemes, grapheme or word pieces~\cite{ji2022asrtest}. 
The following equation demonstrates the goal of an ASR system, which is to find the best textual token sequence $\hat{T}$. This is done by selecting a $T$ from a collection of all tokens, $V$, with the highest probability given a speech sequence $S$.

\begin{equation}\label{asr_background}
\hat{T} = \underset{t\in V}{arg\,max}\,p(T\,|\,S)
\end{equation}

\begin{figure}
\centerline{\includegraphics[width=8cm]{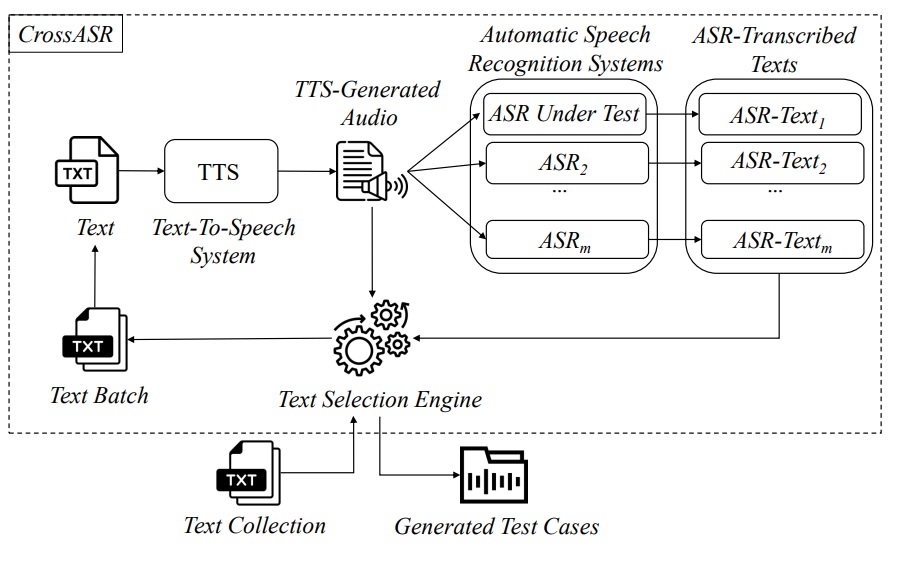}}
\caption{The Architecture of CrossASR~\cite{asyrofi2020crossasr}. It accepts a corpus of texts (\textit{Text Collection}) as input. Each \textit{Text} is fed into the Text-To-Speech System to generate TTS-generated audio. The audio is then processed by ASR system(s), producing \textit{ASR-Transcribed Texts}. These texts are compared with the input \textit{Text} to identify failed test cases.}
\label{fig:CrossASR_Arch}
\end{figure}

\subsection{ASR Testing}
Researchers have proposed different methods for ASR testing, such as CrossASR~\cite{asyrofi2020crossasr}, ASRTest~\cite{ji2022asrtest}, and PROPHET~\cite{yang2023prioritizing}. 
CrossASR~\cite{asyrofi2020crossasr} uses differential testing and does not require manual labelling of data. 
CrossASR uses TTS systems to automatically generate test cases for ASR systems to efficiently build test cases to uncover erroneous behaviour in ASR systems. 
ASRTest~\cite{ji2022asrtest} is an automated testing approach built on metamorphic testing theory. 
The primary objective of ASRTest is to enhance the robustness of ASR systems.
It emphasises the importance of ASR systems capable of accurately transcribing speech audio under diverse conditions, encompassing variations in characteristics, background noise, and acoustic distortions originating from the environment.
ASRTest achieves this goal by incorporating various speech transformation operators designed to assess the robustness of ASR systems. 
These operators include the mutation of speech characteristics, noise injection, and simulation of reverberation. 
Apart from that, PROPHET is a tool that predicts word errors in ASR systems to prioritise test cases that are most likely to be incorrect in order to efficiently uncover more errors for the improvement of ASR systems. 

Given our objective of evaluating false alarms in the context of synthetic test cases, we choose CrossASR~\cite{asyrofi2020crossasr} as the tool in our experiment. 
CrossASR's ability to generate test cases using TTS systems aligns well with our research focus, enabling us to thoroughly assess and analyse false alarms in ASR systems testing.

\subsection{CrossASR}
CrossASR serves the purpose of automating the generation of test cases for ASR systems without the need for manually labelled data. 
This is achieved by using Text-to-Speech (TTS) systems within the CrossASR framework. 
By providing a text input, CrossASR utilises TTS systems to synthesise speech, creating TTS-generated audio for testing.
One of the key techniques employed by CrossASR is differential testing. This approach involves comparing and contrasting the outputs of multiple ASR systems when presented with the same input. 
By analysing and identifying inconsistencies or incorrect behaviours among the outputs of the ASR system, CrossASR effectively detects and highlights areas where ASR systems may be functioning improperly. 

Figure \ref{fig:CrossASR_Arch} shows the architecture of CrossASR. A batch of texts called \textit{Text Batch} is first selected from the \textit{text collection} via the \textit{test selection engine}. The \textit{text batch} contains a number of \textit{text}s. Then, each \textit{text} in \textit{text batch} is fed into \textit{Text-To-Speech system}, which produces a TTS-generated audio of the text. The \textit{TTS-generated audio} is then fed into the ASR systems. ASR systems then transcribe \textit{TTS-generated audio}, producing \textit{ASR-transcribed texts}. The \textit{ASR-transcribed texts} are then compared with the input \textit{text} to identify failed test cases. If the \textit{ASR-transcribed text} produced by the ASR under test matches the \textit{text}, CrossASR determines it as a \textit{successful} test case. If the \textit{ASR-transcribed text} produced by the ASR under test does not match the input \textit{Text} but there is no other \textit{ASR-transcribed text} that matches the input \textit{Text}, CrossASR considers this a \textit{failed} test case. If all the \textit{ASR-transcribed texts} are incorrect, CrossASR considers this as an \textit{indeterminable} case, as there could be an issue with the TTS. 

\section{Methodology} \label{Methodology}
In this section, we present an overview of our preprocessing steps, the methodology used to identify false alarms, and the training process of our false alarm estimator.

\subsection{False Alarm Identification}\label{AA}

\begin{figure}
\centerline{\includegraphics[width=8.5cm]{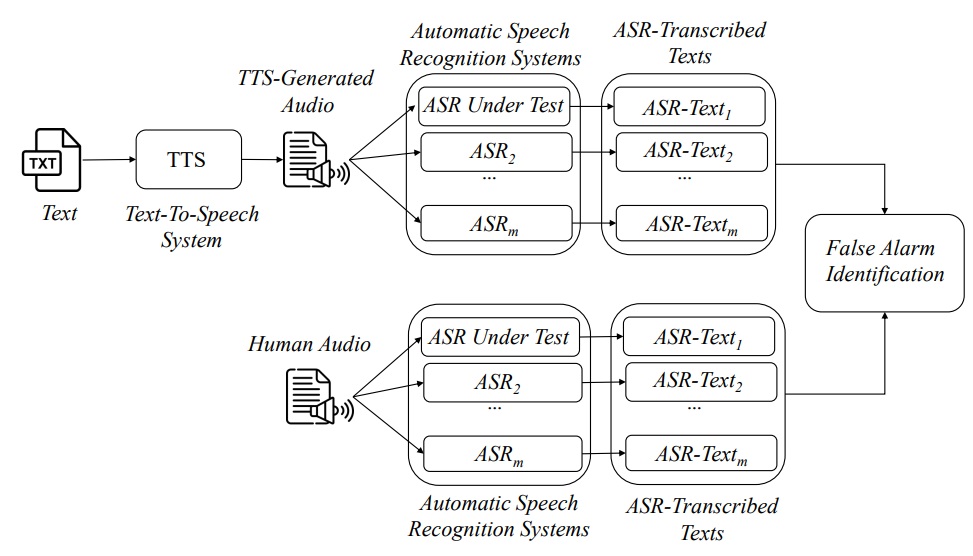}}
\caption{The overview of our proposed methodology to uncover false alarms. }
\label{fig: Arch}
\end{figure}

In our research, a \textit{false alarm} refers to a specific situation where an Automatic Speech Recognition (ASR) system exhibits a failure in transcribing audio that is generated using Text-to-Speech (TTS), while it can accurately transcribe the same content when presented with human audio. 
The inclusion of human audio as a reference is crucial to validate the existence of false alarms when TTS-generated audio is incorrectly transcribed by the ASR system.

Figure~\ref{fig: Arch} shows the overview of our approach to identify false alarms. The input \textit{text} is fed into \textit{Text-To-Speech systems} to produce \textit{TTS-generated audio.} 
The \textit{TTS-generated audio} is fed into \textit{automatic speech recognition systems} to produce \textit{ASR-transcribed texts}. We then provide \textit{human audio} into \textit{automatic speech recognition engines} to produce another set of \textit{ASR transcribed texts}. 
Among the ASR systems used, we specifically choose one for testing purposes, known as the "ASR under test" (as indicated in Figure \ref{fig: Arch}). The \textit{ASR-transcribed texts} of human audio and the \textit{ TTS-generated audio} will then be compared with the input \textit{Text} to determine whether the test case is a false alarm. A test case is deemed a potential false alarm only if the ASR under test accurately transcribes human audio but incorrectly transcribes TTS-generated audio. If the ASR under test flags the test case as a potential false alarm, we will cross-reference with other ASR systems (that is, $ASR_2$, $ASR_m$ in Figure \ref{fig: Arch}) to verify the flagged false alarm. It is considered a true false alarm only if the ASR under test and at least one other ASR fail to correctly transcribe TTS-generated audio. In this study, we assume that the false alarm may be due to TTS-generated audio when more than one ASR fails to transcribe the TTS-generated audio.

In our evaluation process, we use the Word Error Rate (WER) to measure the performance of ASR systems. 
WER is a widely used metric that quantifies the accuracy of ASR by indicating the proportion of incorrectly transcribed words. 
As described by Kepuška et al.~\cite{kepuska2017comparing}, the WER value is proportional to the number of words transcribed erroneously by the ASR system.
The WER is determined by calculating the number of word errors, which encompasses the combined sum of insertions (I), deletions (D) and substitutions (S) of words made by the ASR system compared to the reference transcription.
This sum is then divided by the total number of words in the reference transcription (N), as represented by Equation~(\ref{wer}). 
\begin{equation} \label{wer}
WER = \frac{I+D+S}{N}
\end{equation} 
A higher WER value indicates a larger number of incorrectly transcribed words by the ASR system, reflecting a decrease in overall transcription accuracy. By employing the WER metric, we can effectively assess and compare the performance of different ASR systems in terms of their transcription accuracy.






\subsection{Text and Audio Processing}
\noindent{\textbf{Text Processing.}} 
Prior to being input into the Text-to-Speech (TTS) system, a \textit{text} is subjected to a preprocessing stage. 
The objective of this step is to facilitate accurate pronunciation by standardising the text format.
Similarly, text processing is performed on \textit{ASR-transcribed texts} before \textit{false alarm identification} to standardise the transcriptions. 
This standardisation allows for accurate comparisons when identifying false alarms. Since different ASR systems may transcribe audio in varying ways, slight formatting differences can occur between the \textit{text} and the \textit{ASR-transcribed texts}. 
These discrepancies should not be misconstrued as instances of false alarms, as the intrinsic content remains consistent. 
For instance, an ASR system might interpret the spoken word "one" as the numerical representation "1," but such a difference in form should not be classified as a false alarm, as the inherent meaning is the same.
Specifically, the text processing steps are as follows:

\begin{itemize}[leftmargin=*]
\item Converting all characters to lowercase. 
\item Remove all types of punctuation, except apostrophes.
\item Resolving abbreviations by expanding them to their full form (e.g., “Mr” to “Mister”). 
\item Changing numerals to words (e.g. "1" to "one").
\end{itemize}

\noindent{\textbf{Audio Processing.}}
Before the audio is fed into the ASR systems for transcription, it undergoes a transformation process.
The original audio file is converted to a \texttt{. wav} format, using a sampling rate of 16 kHz and a bit depth of 8. 
This configuration is widely acknowledged as the optimal setting for ensuring maximum speech intelligibility throughout the transcription process~\cite{stan2011romanian}.

\subsection{False Alarm Estimator}


The motivation behind developing a false alarm estimator can be understood through the following scenarios. 
First, the existence of false alarms in the ASR performance evaluation can lead testers to draw incorrect conclusions about the actual performance of the system. 
It also highlights the inefficiency in the testing process, as the resources allocated to identifying these false alarms could have been better used to uncover genuine errors.  
Second, in the development life cycle of an ASR system, failed test cases are analysed and used to improve the ASR system (e.g., by fine-tuning or retraining). However, false alarms do not produce further improvements.
Therefore, integrating an estimator in this scenario allows users to identify potential occurrences of false alarm from the test inputs, help increase the testing efficiency, and facilitate better data collection to improve ASR systems.

Next, we discuss the components of the proposed false alarm estimator. We use the false alarm results produced from our proposed approach presented in Figure~\ref{fig: Arch} as the training dataset and leverage a supervised learning approach to train a text-based false alarm estimator model. 
The false alarm estimator aims to estimate whether a false alarm will arise when a word or sentence is provided as input. 
There are two main reasons for our decision to train a text-based estimator instead of an audio-based one.

One consideration is the simplicity of processing the data. Audio data are typically represented as continuous, time-varying signals, with multiple data points, multiple channels, and variations of amplitude. It is much more complex than text data, which can be represented as a sequence of words. Apart from that, processing audio often requires complex signal processing such as Fourier analysis or spectral analysis. Text data, in comparison, can be processed using simpler techniques such as tokenization or word embedding. 

Another consideration is efficiency. The multidimensional nature and complex temporal structure of audio data require larger and more sophisticated classifiers. As a result, audio-based classifiers usually take longer to train and infer ~\cite{sharma2020trends}. The text-based model proposed in our study can demonstrate better efficiency, allowing us to apply the estimator to large-scale settings.

In this experiment, we used 20 sets of false alarm results obtained from conducting our experiments with four TTS systems (Google, Festival, Espeak, and GlowTTS) and five ASR systems (DeepSpeech, DeepSpeech2, Vosk, Wav2letter++, and Wav2vec2). 
Each text in a dataset is used to generate 20 synthetic test cases, which can potentially be false alarms. 
We count the occurrences of false alarms of each text and rank them according to their total number of occurrences. 
We use the middle number (i.e., 10) as the threshold to prepare examples to train and evaluate our estimator.
More specifically, if a text yields more than 10 false alarms, it is assigned label 1.
Otherwise, a text has the label 0. 
The label quantifies the probability that a text leads to more false alarms when used to generate TTS-generated audio for test ASR systems.
The words in each text in the resulting labelled dataset are then used to build a vocabulary dictionary using the TensorFlow
Tokenizer~\cite{abadi2016tensorflow}. 
The tokenizer assigns an index to each word present in the dataset.
The vocabulary dictionary is sorted by frequency of words. 
Words that have a lower index mean that they appear more frequently in the dataset. 
An example is shown below:

\begin{tcolorbox}[breakable, enhanced jigsaw, rounded corners, parbox=false, boxsep=1pt,left=0.5em,right=0.5em,top=0.5em,bottom=0.5em, title = {An example of vocabulary dictionary}, colback=white]
    \small
   [("the", 1), ("of", 2), ("and", 3), ("to", 4), ("in", 5), ("a", 6), ("was", 7), ("that", 8), ("he", 9), ("his", 10)] \noindent The word "the" has the lowest index, indicating that it appears the most in the dataset.
\end{tcolorbox}

With the vocabulary dictionary ready, each sentence in the dataset is assigned to its numerical form, and each integer represents the index of the word in the vocabulary dictionary. Zeros are padded at the end of each sentence to make all numerical sentence representations of equal length. The following shows an example where each word in the sentence is mapped to its index: "the" is mapped to 1 and "green" is mapped to 614. The dataset with all the numerical representations of the sentences along with their labels is divided into 60\% for model training and 30\% for model testing. 10\% of the training set is treated as the training validation set. 

\begin{tcolorbox}[ enhanced jigsaw, rounded corners, parbox=false, boxsep=1pt,left=0.5em,right=0.5em,top=0.5em,bottom=0.5em, title = {An example of output}, colback=white]
    \small
  \textbf{Input:} the green plant owes its power to absorb the energy of sunlight \noindent 
  \\\
  \textbf{Mapped output:} [1 614 615 7293 55 369 2 4279 1 673 2 2219 0 0 0 0 0 0 0 0 0 0 0 0 0 0 0 0 0 0 0 0 0 0 0 0 0 0 0 0 0 0 0 0 0 0 0 0 0 0 0 0 0 0 0 0 0 0 0 0 0 0 0 0 0 0 0 0 0 0 0 0 0 0 0 0 0 0 0 0 0 0 0 0 0 0 0 0 0
\end{tcolorbox}

A Recurrent Neural Network (RNN) model is built with two layers of Long-Short-Term Memory Units (LSTM)~\cite{hochreiter1997long}. Binary cross-entropy is set as the loss function, and Adam\footnote{Adam is an optimiser algorithm~\cite{zhang2018improved} based on stochastic gradient descent that aims to adjust the learning rate to different parameters.} as the optimiser for model training. We train the model for 20 epochs. Finally, the testing dataset is used as unseen data to evaluate model performance.


\section{Experiments} \label{Experiments}
\subsection{Datasets}
The datasets used in this paper are open-source speech datasets, namely the LJ Speech Dataset~\cite{ljspeech17} and the LibriSpeech Dataset~\cite{panayotov2015librispeech}. These datasets are selected due to their wide usage and popularity in numerous previous studies~\cite{park2020improved, park2019specaugment, li2019jasper}.
LJ Speech Dataset is a collection of short audio clips of a single speaker reading passages from a number of books. It comprises a total of 13,100 English audio recordings, with each recording being no longer than 10 seconds. In total, the dataset provides approximately 24 hours of audio content. 
Each of these recordings, which were recorded from 2016 to 2017 by the LibriVox\footnote{\url{https://librivox.org/}} project, is accompanied by their corresponding transcriptions. After dropping the unusable recordings (e.g. audio without transcription, unclear audio, mismatch between the audio and the corresponding transcription), the LJ Speech Dataset is reduced to 9,925 recordings that are deemed suitable for use in our research. LibriSpeech Dataset is part of the LibriVox project consisting of approximately 1,000 hours of audiobook recordings from Project Gutenberg,\footnote{\url{http://www.gutenberg.org}} a library of free eBooks. Due to hardware and constraints related to computing resources, we have randomly chosen 12,000 recordings with a total duration of approximately 50 hours from the LibriSpeech Dataset for this research. 

\subsection{ASR and TTS Systems}
The ASR systems chosen for this study are: 
Deepspeech~\cite{hannun2014deep}, Deepspeech2~\cite{amodei2016deep}, Vosk~\cite{vosk}, Wav2letter++~\cite{collobert2016wav2letter}, and Wav2vec2~\cite{baevski2020wav2vec}.
Our experiment uses these four ASRs because they are popular and widely used in many studies~\cite{collobert2016wav2letter, hannun2014deep, amodei2016deep, baevski2020wav2vec}. Deepspeech2, developed by Baidu, is widely recognised and used in various related products from Baidu~\cite{baevski2020wav2vec, ji2022asrtest}. Additionally, we select Vosk to represent lightweight models due to its widespread adoption in portable devices. Vosk offers speech recognition capabilities for more than 20 languages and has 5.2k stars on GitHub. 

\begin{figure*}
    \centerline{\includegraphics[scale = 0.65]{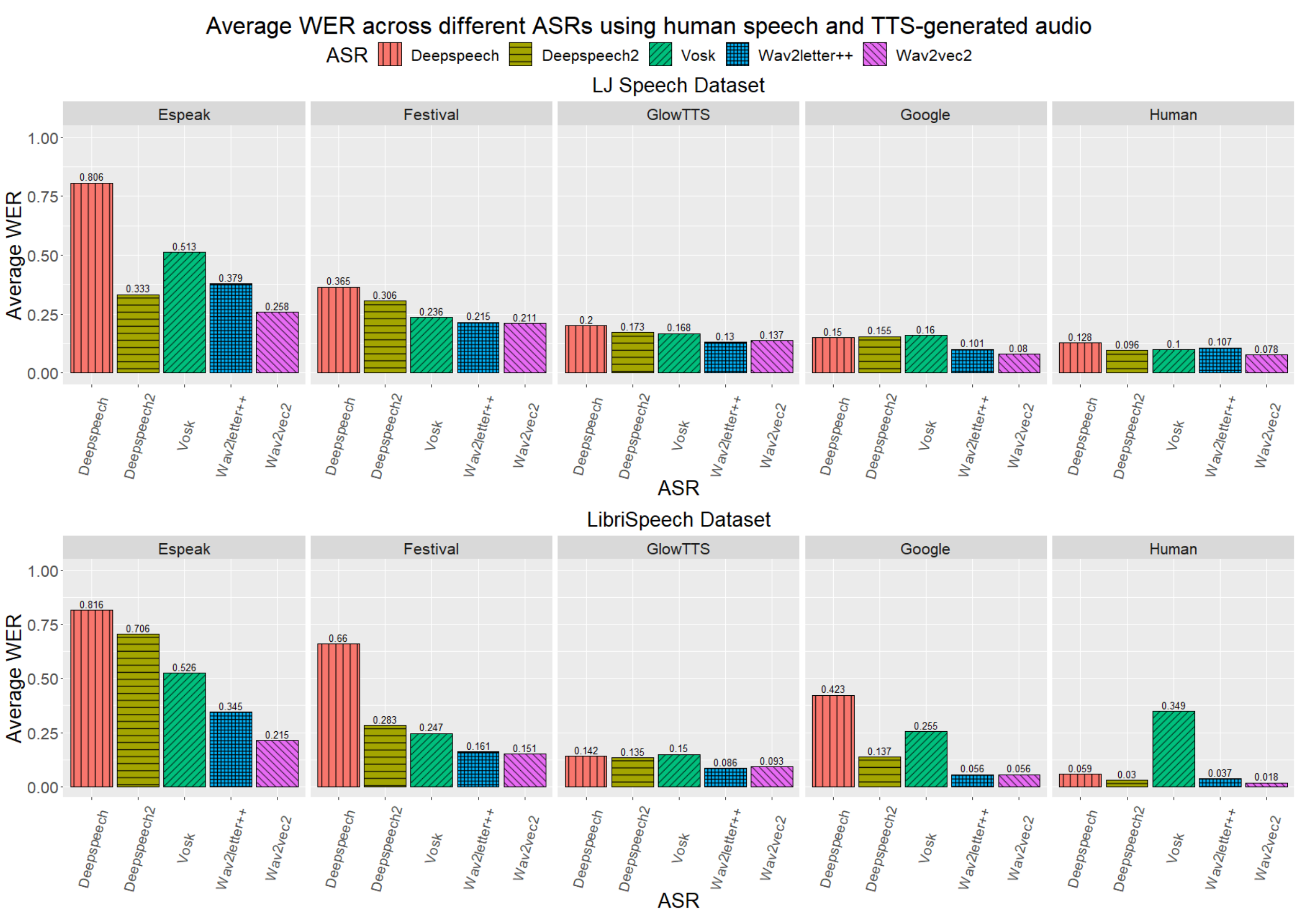}}
    \caption{Average WER for each combination of TTS and ASR systems with human audio and TTS-generated audio.
    \label{fig:WER}}
\end{figure*}

The investigated TTS systems are Google TTS~\cite{googletts}, Espeak~\cite{espeaktts},
Festival~\cite{festival}, and GlowTTS~\cite{glowtts}. In our experiment, we include the first three TTS systems because CrossASR uses them in their study. Additionally, GlowTTS is also introduced into our study to act as a control because it was not originally included in CrossASR. 


\section{Results} \label{Results}

\noindent\textbf{RQ1: How notable is the performance difference between ASR systems when transcribing human audio and \\TTS-generated audio?}

In this research question, our aim is to evaluate the performance variation in each ASR when transcribing TTS-generated and human audio. To achieve this, we run our approach on both the LJ Speech Dataset and the LibriSpeech Dataset. 
We apply our approach to all possible combinations of ASR and TTS systems. 
Following this, we calculate the average Word Error Rate (WER) to quantify the discrepancies in performance across different scenarios.

The average WER for each ASR generated using human audio and TTS-generated audio for each combination of TTS and ASR systems is shown in Figure \ref{fig:WER}. 
The graph is divided into two sections: the top section represents results from the LJ Speech Dataset, while the bottom section corresponds to the LibriSpeech Dataset. 
The vertical axis denotes the average WER and the horizontal axis indicates the different ASR systems. 
In each section, there are five graphs where every graph corresponds to a different audio type: human audio and audio generated by the different TTS systems.

Overall, in both datasets, ASR systems have delivered lower WER values when transcribing human audio than TTS-generated audio. As shown in Figure 3, most ASR systems produce a lower WER when transcribing human audio. On the contrary, these ASR systems exhibit higher WER values when working with TTS-generated audio. This implies that ASR systems generally are better at transcribing human audio than TTS-generated audio. 
On average, ASR systems produce an average WER value of 0.27 when transcribing TTS-generated audio and an average WER value of 0.10 with human audio. There are some cases where an ASR is able to achieve similar low WER results with both TTS-generated audio and human audio. One such example is Wav2Vec2, where it has yielded a WER value of 0.08 and 0.078 on Google-generated audio and human audio, respectively, in the LJ Speech Dataset.


However, there is an exception to this common pattern. Unlike the result in the LJ Speech Dataset shown in Figure \ref{fig:WER}, Vosk performs better on TTS-generated audio than human audio in the LibriSpeech Dataset. This anomaly may arise from the characteristics of Vosk as a lightweight model. It suggests a possible compromise between the model's accuracy and its adaptability for use on lightweight devices, which might be the reason for Vosk's lower accuracy. Even with its relatively lower accuracy, we decided to include Vosk in our study due to its wide adoption in mobile applications. 

In addition to that, there is a clear difference in performance between the ASR systems when transcribing audio produced by different TTS systems. ASR systems have produced high average WER values when transcribing the audio produced by the Espeak TTS system. Most ASR systems have generated low WER values when transcribing GlowTTS-generated audio and Google-generated audio. After manually reviewing the audio produced by the TTS systems, we find that both GlowTTS and Google TTS consistently generate the clearest human-like audio. 
This is based on the manual evaluation by the authors of this work, where we manually sampled 10 audio files generated by each of the TTS systems. Five of the authors evaluated the quality of the sample audio files on a scale of 1 to 5, with 1 being the poorest quality and 5 being the best quality. The average scores given to Espeak, Festival, Google TTS, and Glow TTS are 1.6, 2.2, 4.6, and 4.6 respectively. This shows that Google TTS and GlowTTS tend to produce clear human-like audio files compared to the other TTS systems. As a result, the WER values tend to be lower (better) for both TTS systems. This suggests that the quality of the TTS has an impact on the performance of the ASR.

\begin{tcolorbox}[breakable, enhanced jigsaw, rounded corners, parbox=false, boxsep=1pt,left=0.5em,right=0.5em,top=0.5em,bottom=0.5em]
    \small
    \textbf{Answers for RQ1:} There are notable differences in the performance of ASR systems when transcribing human audio and TTS-generated audio. For LJ Speech Dataset, the average WER for human and TTS-generated audio are 0.102 and 0.254 respectively, while for LibriSpeech Dataset, the average WER are 0.099 and 0.282 respectively.
\end{tcolorbox}

\begin{figure*}[htbp]
    \centerline{\includegraphics[scale=0.6]{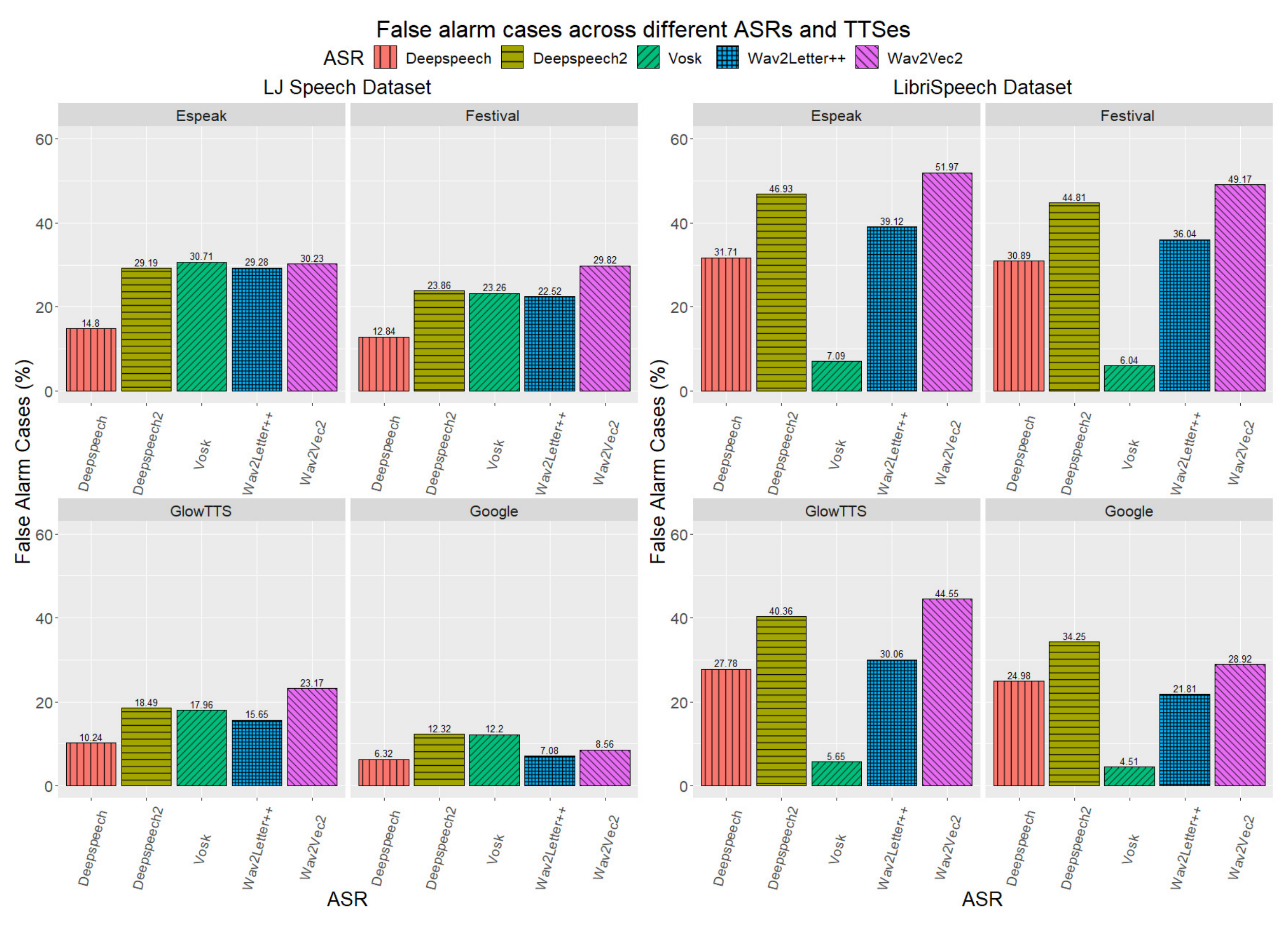}}
    \caption{Percentage of false alarms out of all the test cases for each combination of TTS and ASR systems
    \label{fig:FA_TTS}}
\end{figure*}


\noindent\textbf{RQ2: How prevalent are false alarms when using \\TTS-generated audio to test ASR systems?}

In this research question, we investigate the frequency of false alarms when TTS-generated audio is used to test ASR systems. To do this, we run our proposed pipeline shown in Figure \ref{fig: Arch} for each combination of ASR and TTS systems and observe the number of test cases that are flagged as false alarms. The results are shown in Figure \ref{fig:FA_TTS}, Table \ref{tab:lj_number_fa_produced} and Table \ref{tab:libri_number_fa_produced}. 
Figure \ref{fig:FA_TTS} shows a detailed comparison of the percentages of false alarms from test cases for each combination of ASR and TTS systems with the LJ Speech Dataset and the LibriSpeech Dataset. Table \ref{tab:lj_number_fa_produced} and Table \ref{tab:libri_number_fa_produced} show the total number of false alarms generated for all combinations of TTS and ASR systems with the LJ Speech Dataset and LibriSpeech Dataset respectively. The number of false alarms illustrates the limitations of using TTS-generated audio to test ASR systems. The more false alarms are produced, the more unreliable the TTS-generated audio is for testing ASR systems.

In the LJ Speech Dataset, the number of false alarms for Deepspeech, Deepspeech2, Vosk, Wav2letter++ and Wav2vec2 is 4,386, 8,323, 8,351, 7,397 and 9,110 respectively, as shown in Table \ref{tab:lj_number_fa_produced}. Deepspeech produces the least number of false alarms with all TTS systems, as shown in Figure \ref{fig:FA_TTS} whereas Wav2vec2 produces the most false alarms with most TTS systems. 

In the LibriSpeech Dataset, the number of false alarms for Deepspeech, Deepspeech2, Vosk, Wav2letter++ and Wav2vec2 are 13,844, 19,961, 2,795, 15,244, and 20,852, respectively, as shown in Table \ref{tab:libri_number_fa_produced}. Although Vosk has produced the lowest false alarms with LibriSpeech, it cannot be taken into account due to the result inconsistency shown in Figure \ref{fig:WER}. In Figure \ref{fig:WER}, when tested on the LibriSpeech Dataset, Vosk is shown to have produced a higher WER when transcribing human audio than the audio generated by some of the TTS systems. This means that Vosk is unable to transcribe human audio in LibriSpeech Dataset, and there are a lot of failed test cases for human audio. With Vosk struggling to transcribe human audio from LibriSpeech Dataset, we are unable to validate the occurrences of false alarms with Vosk and thus Vosk is omitted for analysis.
With Vosk excluded, we can observe that Deepspeech produces the least number of false alarms, and Wav2vec2 yields the most false alarms, which is similar to the results from the LJ Speech Dataset. Thus, we can consider Deepspeech as the best ASR in terms of false alarm occurrences. Due to the large amounts of test cases required to test a software system, even a small percentage of false alarms will cost developers a lot of time and effort to resolve. Developers have the expectation that a testing tool should produce fewer than 10\% false alarms to avoid fixing non-existent issues \cite{sadowski2018lessons}, indicating that 10\% should be the threshold of acceptable false alarm rate. Therefore, looking at our results in Figure \ref{fig:FA_TTS}, many of the false alarm rates have exceeded 10\%, which means that false alarms are prevalent when using TTS-generated audio to test state-of-the-art ASR systems.

\begin{table}[!t]
    \setlength{\abovecaptionskip}{-0.1em}
    \setlength{\linewidth}{5em}
    \caption{LJ Speech - Number of False Alarms}
    \begin{center}
    \begin{tabular}{l p{.45\linewidth} p{.45\linewidth} p{.45\linewidth} p{.45\linewidth} p{.45\linewidth} p{.45\linewidth}}
    \toprule
    & \textbf{DS} & \textbf{DS2} & \textbf{VK} & \textbf{W2L} & \textbf{W2V2} & \textbf{Total} \\
    \midrule
    \textbf{Google} & 627 & 1,223 & 1,211 & 703 & 850 & \textbf{4,614} \\
    \textbf{Festival} & 1,274 & 2,368 & 2,309 & 2,235 & 2,960 & \textbf{11,146} \\
    \textbf{Espeak} & 1,469 & 2,897 & 3,048 & 2,906 & 3,000 & \textbf{13,320} \\
    \textbf{GlowTTS} & 1,016 & 1,835 & 1,783 & 1,553 & 2,300 & \textbf{8,487} \\
    \midrule
    \textbf{Total} & 4,386 & 8,323 & 8,351 & 7,397 & 9,110 & \textbf{37,567} \\
    \bottomrule
    \end{tabular}
    \end{center}
    \footnotesize{DS: Deepspeech, DS2: Deepspeech2, VK: Vosk, W2L: Wav2letter++, W2V2: Wav2vec2}
    \label{tab:lj_number_fa_produced}
\end{table}

\begin{table}[!t]
    \setlength{\abovecaptionskip}{-0.25em}
    \setlength{\linewidth}{5em}
    \caption{LibriSpeech - Number of False Alarms}
    \begin{center}
    \begin{tabular}{l p{.45\linewidth} p{.45\linewidth} p{.45\linewidth} p{.45\linewidth} p{.45\linewidth} p{.45\linewidth}}
    \toprule
    \textbf{} & \textbf{DS} & \textbf{DS2} & \textbf{VK} & \textbf{W2L} & \textbf{W2V2} & \textbf{\textbf{Total}} \\
    \midrule
    \textbf{Google} & 2,998 & 4,110 & 541 & 2,617 & 3,470 & \textbf{13,736} \\
    \textbf{Festival} & 3,707 & 5,377 & 725 & 4,325 & 5,900 & \textbf{20,034}\\
    \textbf{Espeak} & 3,805 & 5,631 & 851 & 4,694 & 6,236 & \textbf{21,217} \\
    \textbf{GlowTTS} & 3,334 & 4,843 & 678 & 3,608 & 5,246 & \textbf{17,709} \\
    \midrule
    \textbf{Total} & 13,844 & 19,961 & 2,795 & 15,244 & 20,852 & \textbf{72,696} \\
    \hline
    \end{tabular}
    \end{center}
    \footnotesize{DS: Deepspeech, DS2: Deepspeech2, VK: Vosk, W2L: Wav2letter++, W2V2: Wav2vec2}
    \label{tab:libri_number_fa_produced}
\end{table}

In Research Question 1 and in Figure \ref{fig:WER}, we note that the quality of TTS affects the performance of ASR, with Google TTS and GlowTTS being the best quality among the other TTS systems. In Figure \ref{fig:FA_TTS}, we also observe that there is a correlation between the quality of TTS systems and the number of false alarms. ASR systems have generated the lowest false alarms with Google TTS and the second lowest with GlowTTS. In contrast, all chosen ASR systems produce the most false alarms when transcribing Espeak-generated audio. This is consistent with the WER results in Figure \ref{fig:WER}, and this observation suggests that the quality of the TTS-generated audio plays a crucial role in the occurrence of false alarms. Therefore, using a higher quality TTS will result in a lower WER value, and thus fewer false alarms. The lower the percentage of false alarm occurrence, the more preferable the TTS is for automated testing of ASR systems. 

\begin{tcolorbox}[ enhanced jigsaw, rounded corners, parbox=false, boxsep=1pt,left=0.5em,right=0.5em,top=0.5em,bottom=0.5em]
    \small
    \textbf{Answers for RQ2:} The occurrence of false alarms is prevalent when using TTS-generated audio to test ASR systems. When evaluated on LJ Speech Dataset, only experiments conducted using Google TTS are found with fewer than 10\% false alarms. For all other ASR and TTS systems, on average false alarm cases are found to exceed 15\%. The results also suggest that the prevalence of false alarms depends on the audio quality of TTS-generated audio.
\end{tcolorbox} 

\noindent\textbf{RQ3: How to estimate the potential occurrences of false alarms when testing ASR systems?}

We use the results obtained in RQ2 to train a Recurrent Neural Network (RNN) classifier to estimate whether a test case is likely to lead to false alarms. 
Training and test data are pre-processed as discussed in Section 3.3 to transform the results into a suitable input format for the RNN. 
The dataset containing 4,769 positive and 17,156 negative examples, is randomly shuffled and split into 60\% for model training, 10\% as a validation dataset, and 30\% for model testing, resulting in 13,156, 2,192 and 6,577 cases, respectively. 
The model is trained with 20 epochs. 
Binary cross entropy is used as a loss function for model fitting. 

We use precision, recall, accuracy, and the F1 score for model performance evaluation. 
On the unseen test set, the model achieves a high precision of 98.3\%. 
This indicates that 98.3\% of the estimated positive instances are actual false alarms, showing that there is a low false positive rate. 
The recall is 0.964, indicating that our model correctly estimates 96. 4\% of the actual false alarms in the test data set. 
The F1 score for our model is 97.3\%, demonstrating its ability to estimate true positives (false alarms).

By observing the results estimated by the false alarm estimator, we aim to gain insight into the specific words or linguistic features that could be contributing to the occurrence of false alarms. These words exhibited distinctive properties that can challenge TTS systems in synthesising clearly and misleading ASR systems, resulting in false alarms. A prime example of such words are homophones. Further elaboration on this finding can be found in Section 6.1.


\begin{tcolorbox}[enhanced jigsaw, rounded corners, parbox=false, boxsep=1pt,left=0.5em,right=0.5em,top=0.5em,bottom=0.5em]
    \small
    \textbf{Answers for RQ3:} The evaluation results show that our proposed RNN model is capable of automatically estimating potential false alarms. Our model has shown a precision of 98.3\%, a recall of 96.4\%, an accuracy of 98.5\%, and an F1 score of 97.3\%. The ability to estimate potential occurrences of false alarms can be leveraged as a complementary mechanism to carefully choose test inputs for testing ASR systems.
\end{tcolorbox}

 \section{Discussion} \label{Discussion}
 \subsection{Manual Analysis of False Alarms}
We present examples of false alarms and their patterns observed by manual analysis. The analysis procedure is as follows:

\vspace{0.2cm}
\noindent{\textbf{Data filtering.}} For results of each ASR-TTS combination, we compile a list of words that ASR systems incorrectly transcribe when processing TTS-generated audio. Then, we focus on the top 10 words that are most frequently incorrectly transcribed. From there, we select false alarms where at least half of the cross-reference ASR systems have flagged those cases as false alarms. This is done to ensure that the cases that we analyse affect the majority of ASR systems, and we can compare the incorrect transcriptions to observe a pattern. Given time constraints and labour intensiveness of manual inspection, we analyse a sample size of 500 false alarms, which is larger than the statistically representative sample size computed with a confidence level of 95\% and a confidence interval of 5.

\vspace{0.2cm}
\noindent{\textbf{Pattern finding.}} 
To ensure consistency and reliability in our findings while observing patterns, the manual analysis is conducted by five authors of this paper in the following manner.

\begin{enumerate}
    \item The audio files of the chosen false alarm cases (TTS-generated audio) are listened to manually.
    \item Each author compares the false alarm audio files with the ASR-transcribed texts.
    \item Each author compares the false alarm audio files with the ground truth.
    \item Each author compares the patterns they have observed and eliminates infrequent patterns, that is, patterns that occur less than three times.
\end{enumerate}


From these patterns, we can observe certain characteristics of a TTS and limitations in ASR that can result in incorrect transcriptions, and thus leading to false alarms. As such, we have identified the following characteristics and limitations.

\subsubsection{TTS System's Pronunciation} In our analysis, we frequently find that pronunciation is a major contributing factor to incorrect transcriptions by ASR systems. The following are the most common pronunciation problems:

\hfill\\ 
\noindent\textbf{\textit{TTS cannot pronounce consonants prominently}}
\hfill\\
One of the common pronunciation problems observed in TTS-generated audio is the vague pronunciation of consonants, which often leads to inconsistent transcriptions by ASR systems. For example, some TTS systems fail to pronounce the word "r" prominently. An example of this is "officers". When the "r" in "officers" is not pronounced clearly, it will be transcribed as "offices" by the ASR systems.

\begin{tcolorbox}[enhanced jigsaw, rounded corners, parbox=false, boxsep=1pt,left=0.5em,right=0.5em,top=0.5em,bottom=0.5em, colback=white]
    \small
    \textbf{Original Text:} \\
    committee of cabinet \ul{\textbf{officers}} as our government has become more complex
    \\\\ 
    \textbf{ASR Transcribed Text using TTS-generated audio:} \\
    committee of cabinet \ul{\textbf{offices}} as our government has become more complex
\end{tcolorbox}

\noindent\textbf{\textit{TTS cannot pronounce suffixes clearly}}
\hfill\\
Another issue with TTS is their inability to clearly pronounce suffixes, leading to incorrect transcriptions. Some common examples of suffixes that are mistranscribed by ASR systems are "-ing" and "-ed"/"-d". For example, the "-ed" in "asked" is not pronounced prominently by the TTS, and hence the ASR transcribes it as "ask". 

\begin{tcolorbox}[enhanced jigsaw, rounded corners, parbox=false, boxsep=1pt,left=0.5em,right=0.5em,top=0.5em,bottom=0.5em, colback=white]
    \small
    \textbf{Original Text:} \\
    we \ul{\textbf{asked}} the nation to turn over all its privately held gold dollars for dollars to the government of the united states
    \\\\ 
    \textbf{ASR Transcribed Text using TTS-generated audio:} \\
    we \ul{\textbf{ask}} the nation to turn over all its privately held gold dollars for dollar to the government of the United States
\end{tcolorbox}

\noindent\textbf{\textit{TTS cannot pronounce grammatical words}}
\hfill \\
Grammatical words consist of words such as articles, conjunctions, and pronouns. Several of them are spelt and pronounced quite similarly to each other. If a TTS fails to pronounce these words clearly, the words in the ground truth texts can be mistaken for other grammatical words by ASR systems.
An example of this is the word "as"; if the "a" in "as" is not pronounced prominently, it can lead to the word sounding like "is". 

\begin{tcolorbox}[enhanced jigsaw, rounded corners, parbox=false, boxsep=1pt,left=0.5em,right=0.5em,top=0.5em,bottom=0.5em, colback=white]
    \small
    \textbf{Original Text:} \\
    she identified lee harvey oswald \ul{\textbf{as}} the man who shot the policeman
    \\\\ 
    \textbf{ASR Transcribed Text using TTS-generated audio:} \\
    she identified lee havey oswald \ul{\textbf{is}} the man who shot the policeman
\end{tcolorbox}

\subsubsection{ASR Systems Fail to Transcribe Homophones} Homophones are words that are pronounced similarly but have different definitions and spellings. To illustrate, "brake" and "break" are homophones. This is still an open challenge in ASR systems that have difficulty differentiating between homophones. For instance, "toll" and "tole" are homophones used in the example shown below, where both pronunciations sound similar but the ASR is unable to determine the ground truth. Therefore, further investigation is needed to improve the ability of ASR systems to transcribing homophones. 

\begin{tcolorbox}[enhanced jigsaw, rounded corners, parbox=false, boxsep=1pt,left=0.5em,right=0.5em,top=0.5em,bottom=0.5em, colback=white]
    \small
    \textbf{Original Text:} \\
    and no doubt made the meat also pay \ul{\textbf{toll}}
    \\\\ 
    \textbf{ASR Transcribed Text using TTS-generated audio:} \\
    and no doubt made the meat also pay \ul{\textbf{tole}}
\end{tcolorbox}

\subsubsection{ASR Systems Fail to Transcribe Words with Multiple Pronunciations} Some words may have different pronunciations depending on the speaker's accent. This raises an issue where the ASR cannot transcribe the word appropriately due to the difference in pronunciation. As an example, the name "marley" is pronounced "mar-lei" in human audio, whereas in the TTS it pronounces "Marley" as "mar-li", leading the ASR to transcribe it as "marly". Therefore, further improvements can be made to the ASR’s capability to handle words with multiple pronunciations.

\begin{tcolorbox}[enhanced jigsaw, rounded corners, parbox=false, boxsep=1pt,left=0.5em,right=0.5em,top=0.5em,bottom=0.5em, colback=white]
    \small
    \textbf{Original Text:} \\
    \ul{\textbf{robert marley}} at the time of his arrest called himself a surgical instrument maker 
    \\\\ 
    \textbf{ASR Transcribed Text using TTS-generated audio:} \\
    \ul{\textbf{robert marly}} at the time of his arrest called himself a surgical instrument maker
\end{tcolorbox}

 \section{Related Work}
 \label{Related Work}

\subsection{Testing ASR Systems}
\label{AA}
Researchers have proposed a series of works to evaluate different properties (e.g. robustness~\cite{sensei,catchme}, ethics~\cite{biasfinder,CFSA}, security~\cite{advdoor}) of various AI systems (e.g., code models~\cite{alert,Yefet2020}, reinforcement learning~\cite{acsac2022gong}, image recognition~\cite{yang2022revisiting}). 
Researchers have also recently proposed various methods to evaluate the quality of ASR systems from various aspects.
In this section, we present previous studies on the testing of ASR systems. 

Iwama and Fukuda~\cite{Iwamaicst19} evaluate the basic recognition capability of ASR systems. 
They use a language model to generate test sentences and use an audio converter (i.e. TTS systems) to generate various audio data. 
However, the audio generated using TTS systems may be invalid. To address this concern, Asyrofi et al.~\cite{asyrofi2020crossasr} propose the use of differential testing (CrossASR) to filter potentially invalid failed test cases by cross-referencing the output of different ASR systems.
The intuition is that if none of the ASR systems can correctly recognise TTS-generated audio, it may be due to the reason that this TTS-generated audio itself is of low quality. 
On top of CrossASR~\cite{asyrofi2020crossasr}, Asyrofi et al. further propose CrossASR++~\cite{crossasrpp}, which incorporates more ASR and TTS systems. 
Yuen et al.~\cite{asdf} apply various text transformations to generate more failed test cases.
Recent studies also show that synthesised speech data can also be used to improve the performance of ASR systems~\cite{asyrofi2021can,9414778,fazel2021synthasr,9746217}. 
To better improve find the valuable test cases that can improve ASR systems, Yang et al.~\cite{yang2023prioritizing} design Prophet to prioritise speech test cases using a BERT-based language model.

There is also a line of work on applying transformations to audio to evaluate the robustness of ASR systems. 
Du et al.~\cite{du2018deepcruiser} propose DeepCruiser, whose main objective is to create an automated testing framework for ASR systems based on recurrent neural networks (RNN).
DeepCruiser can be applied to RNN-based ASR systems, while it is not applicable to ASR systems that use the latest transformer-based architecture (e.g., HuBERT~\cite{HuBERT}).
DeepCruiser incorporates eight metamorphic transformations, such as audio speed variation, into the original audio input to generate new audio test inputs. 
Wu and Rajan~\cite{catchme} use frequency masking to transform audio to change the output of ASR systems (i.e., robustness evaluation).
Ji et al.~\cite{ji2022asrtest} propose ASRTest, a tool that uses the metamorphic testing theory.
They implement three families of transformation operators that can simulate practical application scenarios to generate speeches.
Rajan et al.~\cite{aequevox} design aequevox to test the fairness of ASR systems.
They found that ASR systems are more robust in the audio spoken by men when noise is added to the audio.

\subsection{Detecting False Alarms in Software Systems}\label{AA}
Researchers in software engineering have proposed to detect false alarms in various systems.
Herzig and Nagappan ~\cite{herzig2015empirically} propose a classification model that uses association mining rules to discover patterns between false alarms. The association rules are then used to estimate and classify failed test cases as false test alarms. The rationale is that false alarms exhibit specific patterns that can be used to identify false alarms. The features used for their model are the unique identifiers of the test case executions, the unique identifiers of the test case, the identifier of the executed test step, and a binary field that indicates whether the test step passed or failed. Another study presented by Yoon et al.~\cite{yoon2014reducing} proposes a machine learning-based approach to reduce the number of false alarms for automated static analysis tools. Similarly, false alarm patterns are used to train the support vector machine (SVM) model and to classify false alarms using tree-based abstract syntax feature vectors. Although many tools have been proposed to estimate false alarms, those tools mainly focus on software testing, and there is still a gap in the false alarm estimator for ASR testing. In our approach, we have taken inspiration from Herzig and Nagappan"s study~\cite{herzig2015empirically} where the training features we used are an array or field of values where each element represents an index of a word based on a vocabulary dictionary. The word comes from the transcription of both successful and false alarm test cases. We have also taken inspiration from the study by Yoon, Jin, and Jung~\cite{yoon2014reducing} to adopt supervised machine learning for our false alarm estimator. One difference is that we chose to use RNN instead of SVM.

\section{Threats to Validity} \label{Threats}

\noindent \textbf{Internal Validity}. When the ASR under test flags a test case as a false alarm, it may not be a true false alarm, as it can occur due to the limitations of the ASR itself. To mitigate this threat, we cross-reference with other ASR systems to verify the false alarm in the false alarm identification step of our approach. As we have a small number of ASR systems, we consider it a true false alarm only when the ASR under test and another ASR in the pool flag the test case as a false alarm. However, having at least one other ASR failing the test case may not be enough to fully validate the false alarm. Given a scenario in which most cross-referenced ASR systems fail in the test case, we still consider that test case as a false alarm. This would indicate that the false alarm is valid only for the ASR under test and for the minority of cross-referenced ASR systems. Therefore, the false alarm data collected are not fully representative of all ASR systems. Apart from that, for the proposed false alarm estimator, we set a threshold such that if a text yields more than 10 false alarms, it is assigned with the label 1 (flagged as false alarm). This assumption might not hold when the estimator is evaluated with significantly more ASR and TTS. As such, the reported findings for our proposed false alarm estimator can only be confined to the chosen ASR and TTS used in this paper.
In the future, especially for research that involves a larger number of ASR systems, this should be mitigated by only considering the false alarm as valid should the majority or at least half of the total cross-referenced ASR systems flag the test case as a false alarm.

\vspace{0.2cm}
\noindent \textbf{External Validity}. The results for RQ1 and RQ2 depend on the datasets selected for our experiment. As many of our selected ASR systems have been trained with the LibriSpeech dataset~\cite{baevski2020wav2vec, collobert2016wav2letter, amodei2016deep}, it can introduce bias where these ASR systems may perform better with said dataset. 
Despite this, the results observed from these ASR systems do not show consistent high performance. This emphasises the necessity of evaluating these ASRs using the LibriSpeech Dataset to better understand their limitations and strengths. On the other hand, by employing a dataset that contains texts already familiar to the ASR systems, we can more confidently attribute false alarms to synthesised audio from the text (i.e., the quality of the TTS systems). As a step to mitigate bias, we also incorporate the LJSpeech Dataset in our experiments, which is unseen by the ASR systems.
The selection of ASR and TTS systems is crucial in our experiment. The TTS quality and ASR's ability in transcribing TTS-generated audio impact the number of false alarms generated in our experiment. To minimise the potential bias that comes from this, we used five ASR systems and four TTS systems. 

 \section{Conclusion and Future Work} \label{Conclusion}
We analyse the difference in performance between ASR systems when transcribing human audio and TTS-generated audio, while also evaluating the prevalence of false alarms in ASR systems tested with TTS-generated audio. 
We also developed a false alarm estimator to help estimate false alarms. This tool allows developers to efficiently estimate the possible occurrences of false alarms without the need for manual verification of all failed test cases. The trained estimator has a precision of 98.3\% and a recall of 96.4\% when tested with the LJ Speech Dataset and the LibriSpeech Dataset, showing good signs of possible future adoption for efficient ASR testing. 
Although TTS-generated audio presents opportunities for cost-effective and time-effective ASR testing, false alarms remain a limitation. Our study highlights the reasons behind false alarms and discusses strategies for addressing this issue, including developing a false alarm estimator to help developers identify false alarms efficiently. We propose to augment CrossASR with the false alarm estimator to identify potential cases of false alarm test cases. Furthermore, the use of high-quality TTS systems can improve the usability of CrossASR and further optimise ASR testing procedures, eliminating the need for manual verification and the creation of test cases.

In the future, we plan to evaluate our false alarm estimator using other datasets. Our false alarm estimator is primarily trained and tested with results obtained from the LJ Speech Dataset and the LibriSpeech Dataset. 
This suggests that its effectiveness may vary when applied to different datasets. 
To achieve an estimator model that is applicable across different datasets, we believe cross-dataset testing and further fine-tuning must be carried out. 
Furthermore, we would like to explore the use of a speech or multimodal approach to train the proposed false alarm estimator to further enhance the performance and capabilities in estimating potential occurrences of false alarms.

In conclusion, developers should be aware of testing with TTS-generated audio, as false alarms have been proven to affect results. This also recommends that researchers consider the active usage of human audio in research, taking into consideration the cons of doing so, such as poor scalability and expense. Alternatively, incorporating a false alarm estimator can still allow researchers to utilise TTS-generated audio by able to find the likely false alarms. With this in mind, we hope that our research presented in this paper can act as a support for the further development of automated ASR testing systems.

\section*{Acknowledgment}
This research is supported by the Ministry of Education, Singapore under its Academic Research Fund Tier 3 (Award ID: MOET32020-0004). Any opinions, findings and conclusions or recommendations expressed in this material are those of the author(s) and do not reflect the views of the Ministry of Education, Singapore.
\bibliographystyle{ACM-Reference-Format}
\balance
\bibliography{mybibliography}

\end{document}